\documentstyle[12pt]{article}

\catcode`@=11
\def\citer{\@ifnextchar [{\@tempswatrue\@citexr}{\@tempswafalse\@citexr[]}}

\def\@citexr[#1]#2{\if@filesw\immediate\write\@auxout{\string\citation{#2}}\fi
  \def\@citea{}\@cite{\@for\@citeb:=#2\do
    {\@citea\def\@citea{--\penalty\@m}\@ifundefined
       {b@\@citeb}{{\bf ?}\@warning
       {Citation `\@citeb' on page \thepage \space undefined}}%
\hbox{\csname b@\@citeb\endcsname}}}{#1}}
\catcode`@=12

\newcommand{\SM}{\mbox{${\cal SM}$}}
\newcommand{\MSSM}{\mbox{${\cal MSSM}$}}
\newcommand{\CP}{\mbox{${\cal CP}$}}
\newcommand{\HIGLU}{HIGLU}
\newcommand{\TeV}{\mbox{Te$\!$V}}
\newcommand{\GeV}{\mbox{Ge$\!$V}}
\newcommand{\tb}{\mbox{tg$\beta$}}

\newcommand{\nn}{\noindent}

\begin{document}

\begin{titlepage}

\begin{flushright}
Internal Report \\
DESY T--95--05 \\
hep-ph/9510347 \\
October 1995
\end{flushright}

\vspace{1cm}

\begin{center}
\baselineskip25pt

{\large \sc \bf
\HIGLU: \\ A Program for the Calculation of the Total Higgs
Production Cross Section at Hadron Colliders via Gluon Fusion
including QCD Corrections}

\end{center}

\vspace{1cm}

\begin{center}
\baselineskip12pt

{\large \sc Michael Spira}\footnote{e--mail address: spira@desy.de}

\vspace{1cm}

{\it II.~Institut f\"ur Theoretische Physik\footnote{Supported by
Bundesministerium f\"ur Bildung und Forschung (BMBF), Bonn,
Germany, under Contract 05 6 HH 93P (5) and by EU Program
{\it Human Capital and Mobility} through Network {\it Physics
at High Energy Colliders} under Contract CHRX--CT93--0357
(DG12 COMA).}, Universit\"at Hamburg,\\ Luruper Chaussee 149,
22761 Hamburg, Germany}

\end{center}

\vspace{2cm}

\begin{abstract}
A program for the calculation of the total Higgs production cross
section via gluon fusion at hadron colliders including
next--to--leading order QCD corrections is presented.
It is suitable especially for Standard Model Higgs bosons and the
neutral Higgs particles of the minimal supersymmetric extension.
The program provides a model--independent calculation for scalar [\CP--even]
as well as pseudoscalar [\CP--odd] Higgs bosons including the
contributions of virtual top and bottom quarks inside the loop coupled
to the Higgs particles.
The relevant input parameters can be chosen from an input file.
As a special case the minimal supersymmetric extension of the Standard
Model can be investigated.  The corresponding couplings are
implemented including the leading higher order corrections.
\end{abstract}

\end{titlepage}

\section{Introduction}
At the future hadron collider LHC the Higgs boson will be produced
primarily via the gluon fusion mechanism for the entire relevant Higgs
mass range within the Standard Model [\SM] \cite{glufus} as well as the
minimal supersymmetric extension [\MSSM] \cite{glufumssm}. Recently
the next--to--leading order QCD corrections to the production cross
sections of scalar [\CP--even] as well as pseudoscalar [\CP--odd]
Higgs bosons have been calculated \cite{QCDcorr,QCDcorr0}. They are significant
for the theoretical prediction of the cross sections leading to an
increase by up to a factor of two compared to the lowest order
results.

In this paper the program \HIGLU\footnote{Comments or suggestions are
welcome and should be sent to spira@desy.de.} for the calculation of the
total Higgs production cross sections including these next--to--leading order
QCD corrections will be presented. Various relevant input parameters
can be chosen from an input file including a flag specifying the model.
Possible options are the Standard Model, its minimal supersymmetric extension
and a general Higgs model by initializing the Higgs Yukawa
couplings to the heavy quarks appropriately. The program includes
the contribution of the top and bottom quarks in
the loop that generates the Higgs couplings to gluons. Within the Standard
Model as well as in most of the parameter space of the
\MSSM~these contributions provide an excellent
approximation for all cases in practice. Moreover, the program allows to
calculate the
decay widths of Higgs bosons into gluons including next--to--leading
order QCD corrections. The gluonic decay mode plays a significant
r\^ole in the intermediate mass range at future $e^+e^-$ colliders
\cite{QCDcorr}.

The source code of the program is written in FORTRAN. It has been
tested on a VAX--station using the operating system VMS, on
different workstations running under UNIX and IBM computers with
the operating systems CMS and TSO. The numerical
integration is performed by using the VEGAS--package \cite{VEGAS}
for integrals of dimension up to three. Parton distributions can
be attached to the program in any desirable way by adjusting the
corresponding subroutine as described in section \ref{sc:struc}.
As the standard parametrization the program contains the GRV sets \cite{GRV}.

\section{Results}
\subsection{$pp \to \Phi + X$}
The hadron cross section of Higgs boson\footnote{The scalar [$\cal CP$--even]
Higgs particles will generically be denoted by $\cal H$, the pseudoscalar
[$\cal CP$--odd] by $A$ and all the neutral Higgs bosons by $\Phi$.}
production via gluon fusion
$gg\to \Phi$ $(\Phi = {\cal H},A)$ including
next--to--leading order QCD corrections, can be cast into the form
\begin{equation}
\sigma (pp \to \Phi + X) = \sigma^\Phi_{LO}
+ \Delta \sigma^\Phi_{virt} + \Delta \sigma^\Phi_{gg}
+ \Delta \sigma^\Phi_{gq} + \Delta \sigma^\Phi_{q\bar q}
\label{eq:glufus}
\end{equation}
with the lowest order cross sections
\begin{equation}
\sigma (pp \to \Phi + X) = \sigma^\Phi_0 \tau_\Phi
\frac{d{\cal L}^{gg}}{d \tau_\Phi}
\end{equation}
The coefficients are
\begin{eqnarray}
\sigma^{\cal H}_0 = \displaystyle \frac{G_F \alpha_s^2(\mu^2)}{288
\sqrt{2} \pi} \left| \sum_Q g_Q^{\cal H} A^{\cal H}_Q (\tau_Q) \right|^2
& \hspace{2cm} &
\sigma^A_0 = \displaystyle \frac{G_F \alpha_s^2(\mu^2)}{128
\sqrt{2} \pi} \left| \sum_Q g_Q^A A^A_Q (\tau_Q) \right|^2
\end{eqnarray}
They include the Yukawa couplings $g_Q^\Phi$ normalized to the \SM~couplings,
and the quark amplitudes
\begin{equation}
\begin{array}{rclcrcl}
A^{\cal H}_Q (\tau_Q) & = & \displaystyle \frac{3}{2} \tau_Q \left[ 1 +
(1-\tau_Q) f(\tau_Q) \right] & \hspace{1cm} &
A^A_Q (\tau_Q) & = & \tau_Q f(\tau_Q)
\end{array}
\label{eq:ampl}
\end{equation}
The function $f(\tau)$ is defined as
\begin{equation}
f(\tau) = \left\{ \begin{array}{ll}
\displaystyle \arcsin^2 \frac{1}{\sqrt{\tau}} & \tau \ge 1 \\
\displaystyle - \frac{1}{4} \left[ \log \frac{1+\sqrt{1-\tau}}
{1-\sqrt{1-\tau}} - i\pi \right]^2 & \tau < 1
\end{array} \right.
\end{equation}
and the scaling variables are
\begin{equation}
\tau_Q = \frac{4m_Q^2}{m_\Phi^2} \hspace{1cm} \mbox{and}
\hspace{1cm} \tau_\Phi = \frac{m_\Phi^2}{s}
\end{equation}
The parameter $m_Q$ denotes the heavy quark mass, $m_\Phi$ the Higgs boson
mass, $s$ the total center of mass energy squared, $G_F$ the Fermi constant
and $\alpha_s$ the QCD coupling constant. The term $\Delta
\sigma^\Phi_{virt}$ parametrizes the infrared regularized virtual two--loop
corrections and the terms $\Delta \sigma^\Phi_{ij}~\ (i,j = g,q, \bar q)$
the individual collinear regularized real one--loop corrections corresponding
to the subprocesses
\begin{displaymath}
gg \to \Phi g, \hspace{1cm} gq \to \Phi q, \hspace{1cm} q\bar q \to \Phi g
\end{displaymath}
The expressions for $\Delta \sigma^\Phi_{virt}$ and $\Delta \sigma^\Phi_{ij}$
can be found in Refs.\cite{QCDcorr,QCDcorr0}. The gluon luminosity is defined
by
\begin{equation}
\frac{d {\cal L}^{gg}}{d\tau} = \int_\tau^1 \frac{dx}{x} g(x,Q^2)
g(\tau/x,Q^2)
\end{equation}
where $g(x,Q^2)$ denotes the gluon density.
The natural values to be chosen for the renormalization scale $\mu$ of
the strong coupling $\alpha_s (\mu^2)$ and the factorization scale $Q$
of the parton densities is given by the Higgs mass $m_\Phi$.

The program \HIGLU~calculates the five terms in eq.(\ref{eq:glufus})
contributing to the total cross section separately as well as their sum for
all kinds of neutral Higgs bosons $\Phi$. The computation of differential
cross sections can be found in Refs.\cite{dsigma} and is not implemented
in the program.

\section{$\Phi \to gg$}
The decay widths of Higgs bosons $\Phi$ into gluons up to next--to--leading
order are given by
\begin{eqnarray}
\Gamma (\Phi \to gg(g), gq\bar q) & = & \Gamma_{LO} (\Phi \to gg) \left[ 1 +
E_\Phi \frac{\alpha_s}{\pi} \right] \nonumber \\ \nonumber \\
E_\Phi & = & E_{virt}^\Phi + E_{ggg}^\Phi + N_F E_{gq\bar q}^\Phi
\end{eqnarray}
with the leading--order expressions
\begin{eqnarray}
\Gamma_{LO}({\cal H}\to gg) & = & \frac{G_F \alpha_s^2(\mu^2)}{36\sqrt{2}\pi^3}
\left| \sum_Q g_Q^{\cal H} A_Q^{\cal H} (\tau_Q) \right|^2 \nonumber \\
\nonumber \\
\Gamma_{LO} (A \to gg) & = & \frac{G_F \alpha_s^2(\mu^2)}{16\sqrt{2}\pi^3}
\left| \sum_Q g_Q^A A_Q^A (\tau_Q) \right|^2
\end{eqnarray}
The amplitudes $A_Q^\Phi (\tau_Q)$ are defined in eq.(\ref{eq:ampl}).  The
coefficient $E_{virt}^\Phi$ denotes the infrared regularized virtual two--loop
corrections, $E_{ggg}^\Phi$ and $E_{gq\bar q}^\Phi$ the collinear regularized
real one--loop corrections.  The analytical formulae of these contributions
can be found
in Ref.\cite{QCDcorr}.  The parameter $N_F$ fixes the number of light external
flavors produced in the decay $\Phi\to gq\bar q$, which is defined to be
equal to the number of flavors contributing to the QCD $\beta$ function. This
definition maps large logarithms into the running strong coupling
$\alpha_s (\mu^2)$. The natural renormalization
scale $\mu$ of the strong coupling $\alpha_s (\mu^2)$ is given by the
corresponding Higgs boson mass $m_\Phi$.

\section{Input Parameters}
In addition to the source code of the program \HIGLU~an input file defined as
unit 98 is needed, from which the program reads the input parameters.
The name of this input file can be defined in the first OPEN statement
of \HIGLU. It should be noted that the input numbers must {\it not} start
before the equality signs in each corresponding line. The input file contains
the following parameters:

\begin{description}
\item \underline{process:} integer \\
choose the process to be calculated: \\
0: gluon fusion $gg\to \Phi$ \\
1: gluonic decay $\Phi \to gg$

\item \underline{collider:} integer \\
choose the hadron collider mode for the gluon fusion process: \\
0: $pp$ \\
1: $p\bar p$ \\
This flag is only relevant for the gluon fusion process.

\item \underline{energy:} double precision \\
center of mass energy [in \TeV] of the hadron collider for the gluon fusion
process

\item \underline{model:} integer \\
choose the model, in which the process should be calculated: \\
\begin{tabular}{rl}
0: & \SM \\
1: & \hfill \begin{minipage}[t]{14cm}
\MSSM~including the dominant two--loop corrections to the Higgs masses and
couplings
\end{minipage} \\[0.6cm]
2: & \hfill \begin{minipage}[t]{14cm}
\MSSM~including only the the leading one--loop corrections, which increase
as the fourth power of the top mass.
\end{minipage} \\[0.6cm]
3: & any model
\end{tabular}

All stop mixing parameters of the \MSSM~are set equal to zero.

\item \underline{tanbeta:} double precision \\
the \MSSM~parameter \tb, which is irrelevant if the flag model equals zero
or three.

\item \underline{$g_{b,t}$:} double precision \\
the Yukawa couplings of the corresponding Higgs boson to the top and bottom
quarks normalized to the \SM~couplings.  These couplings are only
relevant, if the flag for the model is chosen to be three.
If the calculation is performed within the \MSSM, these parameters are
irrelevant, because the Yukawa couplings are calculated from \tb~and the
chosen Higgs mass $m_{Higgs}$. In the \SM~[flag model = 0] these couplings are
automatically set equal to unity.

\item \underline{$m_{b,t}$:} double precision \\
pole masses of the bottom and top quarks [in \GeV]. Useful values for
these masses are given in the sample of the input file in the appendix:
$m_t = 176~\GeV$, $m_b = 5~\GeV$.

\item \underline{type:} integer \\
choose the neutral Higgs boson type: \\
1: heavy scalar $H$ \\
2: pseudoscalar $A$ \\
3: light scalar $h$ \\
In the \SM~ this flag is set equal to unity automatically.
If the model is chosen to be the \MSSM, all three types of Higgs
bosons can be chosen. If the flag for the model is set equal to three, only the
values 1 and 2 are possible, because the nature of the Higgs boson will be
characterized by its mass and its Yukawa couplings $g_{b,t}$ to the heavy
quarks.

\item \underline{$m_{Higgs}$:} double precision \\
the mass [in \GeV] of the chosen Higgs boson type.

\item \underline{loop:} integer \\
choose one--loop or two--loop formula for the strong coupling $\alpha_s$ in
the $\overline{MS}$ scheme: \\
1: one--loop \\
2: two--loop

\item \underline{choice:} integer \\
choose the input value of the strong coupling $\alpha_s$: \\
1: $\alpha_s$ is defined by the value for $\alpha_s (M_Z)$ \\
2: $\alpha_s$ is fixed by the QCD scale $\Lambda_{\overline{MS}}^{(N_F)}$

\item \underline{alpha$_S$:} double precision \\
strong coupling $\alpha_s$ at scale of the $Z$ boson mass $M_Z$. This parameter
is relevant for the flag choice = 1.

\item \underline{$N_F$:} integer \\
the number of flavors, for which the QCD scale
$\Lambda_{\overline{MS}}^{(N_F)}$ is given. This parameter can be chosen to be
3,4,5 or 6. The QCD scales for different numbers of flavors are computed by
using the quark masses $m_{b,t}$ from the input file and $m_c = 1.5~\GeV$ via
the matching conditions of the $\overline{MS}$ scheme at the thresholds.

\item \underline{lambda:} double precision \\
QCD scale $\Lambda_{\overline{MS}}^{(N_F)}$ in \GeV.

The parameters $N_F$ and lambda are relevant for the flag choice = 2.

\item \underline{$n_{ext}$:} integer \\
number of light external flavors to be included in the gluonic decay width of
the Higgs boson. If $n_{ext}$ is set equal to 4, the bottom contribution $\Phi
\to b\bar b g$ is subtracted, and if $n_{ext}$ equals three, the part $\Phi
\to c\bar c g$ is removed. This parameter can be chosen to be 3,4 or 5.
The running strong coupling $\alpha_s(\mu^2)$ uses the {\it same} number of
flavors.

\item \underline{$mu_{1,2}$:} double precision \\
parameters defining the renormalization scale $\mu$ [in \GeV] of the strong
coupling constant $\alpha_s$ in the following way: \\
\begin{displaymath}
\mu = \mu_1~m_{Higgs} + \mu_2
\end{displaymath}
The variable $\mu_1$ is a dimensionless coefficient of the Higgs mass
$m_{Higgs}$, and $\mu_2$ denotes a fixed scale [in \GeV].

\item \underline{$Q_{1,2}$:} double precision \\
parameters fixing the factorization scale $Q$ [in \GeV] for the gluon fusion
process: \\
\begin{displaymath}
Q = Q_1~m_{Higgs} + Q_2
\end{displaymath}
$Q_1$ denotes a dimensionless coefficient of the Higgs mass $m_{Higgs}$,
whereas $Q_2$ defines a fixed scale [in \GeV].

\item \underline{abserr:} double precision \\
absolute error for the VEGAS integration.

\item \underline{points:} integer \\
number of points for the VEGAS integration. 10000 points yield a sufficient
precision, whereas the results using 1000 points with 5 iterations are
already reliable at the percent level.

\item \underline{itmax:} integer \\
maximal number of iterations for the VEGAS integration.

\item \underline{print:} integer \\
flag for the print out style of the VEGAS iterations: \\
0: no print out \\
1: pretty print out \\
10: print out as a table

\end{description}
A sample of the input file is given in the appendix.

\section{Structure Functions \label{sc:struc}}
For the implementation of structure functions the subroutine STRUC
has to be changed appropriately. This part of the program reads as follows:

\begin{verbatim}
      SUBROUTINE STRUC(X,Q,PDF)
      IMPLICIT DOUBLE PRECISION (A-H,O-Z)
      DIMENSION PDF(-6:6)
      COMMON/FACSC/ISCHEME
C...X          - BJORKEN X
C...Q          - MOMENTUM SCALE  (IN GeV)
C...PDF(-6:6)  - MATRIX CONTAINING  X*P(X,Q)
C...    IPDF = -6 ,  -5 ,  -4 ,  -3 ,  -2 ,  -1 ,0 ,1,2,3,4,5,6
C...         T_BAR,B_BAR,C_BAR,S_BAR,U_BAR,D_BAR,GL,D,U,S,C,B,T

C--- CHOOSE PROTON STRUCTURE FUNCTIONS AND THEIR FACTORIZATION SCHEME
C--- ISCHEME:  0            1
C---           MSBAR        DIS
      ISCHEME=0

      Q2=Q**2
      ISET=2
      CALL PDGRV(ISET,X,Q2,PDF)

      RETURN
      END
\end{verbatim}
The line calling the subroutine PDGRV has to be changed, if another
parametrization should be used. The flag ISCHEME has to be specified
according to the factorization scheme, which has been used for the
parametrization of the parton densities. Note that the array PDF has to be
generated in accordance with the given convention.

\section{Output}
The output of the program \HIGLU~is written to a file unit 99, which contains
the chosen input parameters as well as all results obtained by VEGAS
integrations.  The VEGAS iterations are written to the standard output, if
the flag print is
set equal to 1 or 10. The integrated results contain all individual
contributions of the QCD corrections separately and their total sum. A sample
of the output is presented in the appendix, which can be obtained with the
input file shown before, if the flag model and the Higgs mass $m_{Higgs}$
are specified appropriatly. If
the \MSSM~is chosen as the Higgs model, the pseudoscalar Higgs mass $m_A$
corresponding to the scalar Higgs masses $m_{\cal H}$ is also typed out.

\vspace*{2cm}

\nn
{\large \bf Acknowledgements.}
It is a pleasure to thank A.~Djouadi, D.~Graudenz and P.M.~Zerwas for their
fruitful collaboration in the analysis of the QCD corrections to the Higgs
coupling to gluons, out of which this program has been built--up. Special
thanks go to A.~Djouadi and P.M.~Zerwas for a critical reading of the
manuscript and useful suggestions for the program.

\newpage

\section{Appendix}

\subsection{Input File}
An example of the input file is given by

\begin{verbatim}

PROCESS:  0 = GG --> H         1 = H --> GG
=======

PROCESS  = 0
---------------------------------------------------------------
COLLIDER:   0 = P P              1 = P PBAR
========

COLLIDER = 0
---------------------------------------------------------------
TOTAL ENERGY: [TEV]
=============

ENERGY   = 14.D0
---------------------------------------------------------------
MODEL: 0 = SM  1 = MSSM (TWO-LOOP)  2 = MSSM (ONE-LOOP) 3 = ANY
======

MODEL    = 0
---------------------------------------------------------------
TAN(BETA): (MSSM)
==========

TANBETA  = 1.5D0
---------------------------------------------------------------
COUPLINGS:  G_B = BOTTOM    G_T = TOP
==========  (MODEL = 0)

G_B      = 1.D0
G_T      = 1.D0
---------------------------------------------------------------
QUARK MASSES: [GEV]
=============

M_B      = 5.D0
M_T      = 176.D0
---------------------------------------------------------------
HIGGS TYPE AND MASS [GEV]:  1 = HEAVY SCALAR   2 = PSEUDOSCALAR
==========================  3 = LIGHT SCALAR

TYPE     = 1
M_HIGGS  = 200.D0
---------------------------------------------------------------
SCALES: [GEV] MU = MU_1*M_HIGGS + MU_2:   RENORMALIZATION SCALE
=======        Q =  Q_1*M_HIGGS +  Q_2:     FACTORIZATION SCALE

MU_1     = 1.D0
MU_2     = 0.D0
Q_1      = 1.D0
Q_2      = 0.D0
---------------------------------------------------------------
ORDER OF ALPHA_S:  1 = LO     2 = NLO
=================

LOOP     = 2
---------------------------------------------------------------
DEFINITION OF ALPHA_S:  1 = ALPHA_S (M_Z)   2 = BY LAMBDA (N_F)
======================

CHOICE   = 1
---------------------------------------------------------------
ALPHA_S (M_Z):
==============

ALPHA_S  = 0.118D0
---------------------------------------------------------------
LAMBDA_NF: [GEV] (QCD SCALE)
==========

N_F      = 5
LAMBDA   = 0.226D0
---------------------------------------------------------------
NUMBER OF EXTERNAL LIGHT FLAVORS: (FOR H --> GG)
=================================

N_EXT    = 3
---------------------------------------------------------------
VEGAS:       ABSERR = ABSOLUTE ERROR
======       POINTS = NUMBER OF CALLS
             ITMAX  = NUMBER OF ITERATIONS
             PRINT  = PRINT OPTION FOR INTERMEDIATE VEGAS-OUTPUT
                      0            1             10
                      NO OUPUT     PRETTYPRINT   TABLE

ABSERR   = 0.D0
POINTS   = 10000
ITMAX    = 5
PRINT    = 10
---------------------------------------------------------------

\end{verbatim}

\subsection{Output}

\subsubsection{Gluon Fusion}

\begin{verbatim}

 VEGAS:
 ======
 ABSERR    =    0.000000E+00
 POINTS    =      10000          ITERATIONS   =     5

 GLUON FUSION: GG --> HIGGS
 ==========================

 P P COLLIDER
 ============
 ENERGY    =     14.0000     TEV

 LAMBDA_5  =    0.226232     GEV     NLO-ALPHA_S (M_Z) =    0.118000
 REN-SCALE =     200.000     GEV     FAC-SCALE =     200.000     GEV

 T-MASS    =     176.000     GEV     B-MASS    =     5.00000     GEV

 HIGGS     = H
 ============================================================
 M_H       =     200.000      GEV
 G^H_B     =     1.00000      G^H_T    =     1.00000
 SIG_LO    =     6.42484     +-    0.681792E-05  PB
 SIG_VIRT  =     3.45509     +-    0.366648E-05  PB
 SIG_GG    =     4.78077     +-    0.466710E-02  PB
 SIG_GQ    =   -0.268282     +-    0.709725E-03  PB
 SIG_QQ    =    0.182952E-01 +-    0.173479E-04  PB
 SIG_NLO   =     14.4107     +-    0.472080E-02  PB
\end{verbatim}

\newpage

\subsubsection{Gluonic Decay}

\begin{verbatim}

 VEGAS:
 ======
 ABSERR    =    0.000000E+00
 POINTS    =      10000          ITERATIONS   =     5

 HIGGS --> GG
 ============

 NF_EXT    =     3

 LAMBDA_5  =    0.226232     GEV     NLO-ALPHA_S (M_Z) =    0.118000
 REN-SCALE =     500.000     GEV

 T-MASS    =     176.000     GEV     B-MASS    =     5.00000     GEV

 MSSM (2-LOOP):  TG(BETA) =  1.50
 ==============
  Z-MASS =   91.187     GEV           W-MASS  =   80.330     GEV
  NO MIXING                        SUSY-SCALE =   1000.0     GEV

 HIGGS     = H
 ============================================================
 M_H       =     500.000      GEV
 M_A       =     490.788      GEV
 G^H_B     =     1.47436      G^H_T    =   -0.691622
 GAM_LO    =     9.15160      MEV
 E_VIRT    =     15.2434
 E_GGG     =     7.68668     +-    0.812334E-04
 E_GQQ     =   -0.972425     +-    0.307239E-04
 E_TOT     =     20.0128     +-    0.122860E-03
 GAM_NLO   =     13.9932     +-    0.297226E-04  MEV

\end{verbatim}

\end{document}